\documentclass[aps,prd,twocolumn,showpacs,groupedaddress,floatfix]{revtex4}

\usepackage{graphicx}
\DeclareGraphicsExtensions{.eps, .eps, .jpg, .png}

\def\by{{\bar{y}}}
\def\ba{{a\left(\by\right)}}

\def\be{\begin{equation}}
\def\ee{\end{equation}}
\def\ba{\begin{eqnarray}}
\def\ea{\end{eqnarray}}

\def\bdm{\begin{displaymath}}
\def\edm{\end{displaymath}}
\def\la{~\mbox{\raisebox{-.6ex}{$\stackrel{<}{\sim}$}}~}
\def\ga{~\mbox{\raisebox{-.6ex}{$\stackrel{>}{\sim}$}}~}
\def\bq{\begin{quote}}
\def\eq{\end{quote}}

 at 10truept

\newcommand{\beq}{\begin{equation}}
\newcommand{\eeq}{\end{equation}}
\newcommand{\bea}{\begin{eqnarray}}
\newcommand{\eea}{\end{eqnarray}}
\newcommand{\beqa}{\begin{eqnarray}}
\newcommand{\eeqa}{\end{eqnarray}}

\def\la{~\mbox{\raisebox{-.6ex}{$\stackrel{<}{\sim}$}}~}
\def\ga{~\mbox{\raisebox{-.6ex}{$\stackrel{>}{\sim}$}}~}

\def\ltap{\ \raise.3ex\hbox{$<$\kern-.75em\lower1ex\hbox{$\sim$}}\ }
\def\gtap{\ \raise.3ex\hbox{$>$\kern-.75em\lower1ex\hbox{$\sim$}}\ }
\def\gl{\ \raise.5ex\hbox{$>$}\kern-.8em\lower.5ex\hbox{$<$}\ }
\def\roughly#1{\raise.3ex\hbox{$#1$\kern-.75em\lower1ex\hbox{$\sim$}}}

\begin{document}



\title{A Natural Framework for Chaotic Inflation}


\author{Nemanja Kaloper}
\affiliation{Department of Physics, University of
California, Davis, CA 95616, USA}

\author{Lorenzo Sorbo}
\affiliation{Department of Physics,
University of Massachusetts, Amherst, MA 01003, USA}



\begin{abstract}

We show that inflation with a quadratic potential occurs naturally in theories where an axion-like field mixes with a 4-form. Such an axion is massive, with the mass which arises from the mixing being protected by the axion shift symmetry. The 4-form backgrounds break this symmetry spontaneously and comprise a mini-landscape, where their fluxes can change by emission of membranes. Inflation can begin when the 4-form dominates the energy density. Eventually this energy is reduced by membrane emission and the axion can roll slowly towards its minimum, as in the simplest version of chaotic inflation.

\end{abstract}


\maketitle


Among the many scenarios of inflation, the one which stands out in terms of its simplicity, elegance and phenomenological success is chaotic inflation \cite{chaotic}. It has since become a prototype of slow roll inflation, arising as an effective description in many complicated models of inflation. It also fits the observational data really well \cite{wmap}. For the scenario to work, however, one needs the inflaton to initially have super-Planckian expectation values, $\phi \gg M_{Pl}$, in order for the slow roll conditions to be met for long enough, and yield at least $\sim 65$ efolds of inflation. This issue has led to considerable debate, starting with \cite{enqvist}, about how realistic it is to model the inflaton potential by a quadratic term when $\phi \gg M_{Pl}$. In this regime, higher-order corrections to the effective potential might become important, and steepen the potential, spoiling slow roll conditions, or even altogether obstructing inflation. This does not always happen. E.g., the
higher-dimension operators in the loop expansion of the effective potential may seem individually dangerous, going as $(\phi/M)^n$ for some $UV$ scale $M \la M_{Pl}$. However they come in as an alternating series, and when inflaton has power-law self-interactions they sum up to logarithmic corrections, as in the Coleman-Weinberg theory \cite{cw}. Further, the worries of \cite{enqvist} that graviton loops can yield large corrections to the potential do not materialize \cite{andreirad} because graviton one-loop corrections yield terms like $\partial^2_\phi V R$ and ${V^2}/M_{Pl}^4$, which are small where chaotic inflation is presumed to operate, $\partial^2_\phi V  < M^2_{Pl}$ and $V < M^4_{Pl}$. In fact, a simple argument can be fielded explaining how the potential may remain flat despite radiative corrections. If the potential were exactly flat the theory would have had a shift symmetry $\phi \rightarrow \phi + \phi_0$. Radiative corrections would not break it, and the full effective potential would only involve $\partial_\mu \phi$. Conversely, if the potential depends weakly on $\phi$, this shift symmetry is softly broken, and so the radiative corrections are proportional only to the symmetry breaking terms. If the symmetry breaking terms are small, the radiative corrections will stay under control, keeping the effective potential sufficiently flat.

Hence the task is to find theories where dynamics which gives mass to the inflaton is radiatively stable. If so, the inflaton mass and other polynomial interactions will be small enough that further corrections may not spoil their flatness, as per the argument above. This makes various pseudo-Nambu-Goldstone bosons \cite{natural} obvious inflaton candidates (we will call them `axions' henceforth), because their masses arise from non-perturbative effects, whereas the perturbative shift symmetry prevents large radiative corrections~\footnote{Alternative ideas for supergravity inflation were elaborated in~\cite{GMO}.}. The effective potential arises from instanton effects, and can be written in the form of a `Fourier series', 
$\sum_{n} \lambda_{n}{}^4 \cos(2n\phi/f_\phi)$, where $f_\phi$ is the axion decay constant, and $\lambda_n$ are dynamically generated scales in the instanton expansion, typically related to the $UV$ cutoff via $\lambda_1 \sim M e^{-\alpha/g}$, where $g$ is the gauge coupling and 
$\alpha$ a dimensionless number, and with $\lambda_{n>1} < \lambda_1$ (see, e.g. \cite{tometc,nflat}). For gravitational instantons, these formulae change to $\lambda_n\sim M e^{-n\,M_{Pl}/f_\phi}$. The axion varies over the interval $(0, \pi f_\phi)$. To have slow roll inflation, one needs to have the regime where 
$\phi \gg M_{Pl}$, otherwise the field potential will not dominate the evolution for long enough. These 
requirements beg for $f_\phi \gg M_{Pl}$.
On the other hand, it appears to be difficult to obtain large axion decay constants obeying $f_\phi \ga M_{Pl}$ in $UV$ complete theories \cite{tometc}. So if $f_\phi \la M_{Pl}$ the higher order instanton effects come into play, interfering with inflation with large $\phi \gg M_{Pl}$. To date, the proposals which were devised to address this issue employ either the many 
fields~\cite{nflat} or nontrivial compactifications in string theory~\cite{renata,eva}.

In this Letter, we outline a different framework circumventing this problem. It is a higher energy variant of the dynamics of quintessence which we have discussed recently \cite{ksQfin}. If an axion field mixes with a $4$-form in $4D$ by a bilinear term, it becomes massive, with the mass term which {\it preserves} the axion 
shift symmetry of the action. The shift symmetry only breaks spontaneously after picking the background $4$-form solution \cite{giavilen}. 
Thus the mass is protected from field theory radiative corrections, and the potential can only be slanted by instanton effects. Such effects are inevitable, since in order for the axion to be an inflaton, it must have matter couplings so that it can reheat the universe after the end of inflation. 
However, if the axion does not couple to any sector which is strongly 
coupled at or above the scale of inflation, the instanton potential terms will be negligible compared to the leading term induced by the $4$-form mixing. 
Inflation can then unravel precisely as described originally in the simple chaotic inflation scenario of Linde \cite{chaotic}, and reheating can proceed by the production of the gauge bosons to which the axion couples directly.

As in \cite{ksQfin} we consider an axion mixing with a $4$-form, via a term $\sim \phi\, \epsilon^{\mu\nu\lambda\sigma} 
F_{\mu\nu\lambda\sigma}$. The action including minimal coupling to gravity has two parts, describing bulk theory and terms describing membrane nucleation dynamics. Without axion-form mixing, such theories have been studied in the context of cosmological constant relaxation \cite{andreipr}-\cite{duff}. When the mixing is turned on, the bulk term is
\begin{eqnarray}
{\cal S}_{bulk} &=&\int d^4 x \sqrt{g} \left( \frac{M_{Pl}^2}{2} R -\frac12({\nabla}\phi)^2 - 
\frac{1}{48} F_{\mu\nu\lambda\sigma}^2 + \right. \nonumber\\
&+&\left. \frac{\mu}{24}\,\phi\, \frac{\epsilon^{\mu\nu\lambda\sigma}}{\sqrt{g}}
F_{\mu\nu\lambda\sigma} + \ldots \right) \, .
\label{actionbulk}
\end{eqnarray}
The ellipsis denote the matter sector contributions, $\epsilon^{\mu\nu\lambda\sigma}$ is the Levi-Civita tensor density, as indicated by the metric determinant terms and $F_{\mu\nu\lambda\sigma}$ is the antisymmetric derivative of the $3$-form potential,  
$F_{\mu\nu\lambda\sigma} = 4 \partial_{[\mu} A_{\nu\lambda\sigma]}$. The parameter
$\mu$ has dimension of mass, as required to correctly normalize the bilinear  $\phi \epsilon^{\mu\nu\lambda\sigma}  F_{\mu\nu\lambda\sigma}$. For now we view it merely as a given parameter, noting that it can arise from either spontaneous breaking of $Z_{2n}$ discrete symmetries \cite{giavilen}, or from dimensional reductions of higher rank form fields \cite{ksQfin,sugra}, as the flux through compact dimensions. The membrane action over its woldvolume $\xi^a$ with induced metric $\gamma_{ab}$ is 
\be
{\cal S}_{brane} \ni \frac{e}{6} \int d^3 \xi \sqrt{\gamma} e^{abc} \partial_a x^\mu  \partial_b x^\nu  \partial_c x^\lambda A_{\mu\nu\lambda} \, ,
\label{actionbrane}
\ee
where the membrane charge $e$ is normalized to the membrane tension. To correctly covariantize it, we must also include the Gibbons-Hawking term for gravity, and its analogue for the $4$-form \cite{duff,ksQfin}, which are 
$\int d^4 x \sqrt{g} \, \nabla_\mu ( F^{\mu\nu\lambda\sigma}\, A_{\nu\lambda\sigma})/6$  and
$- \int d^4 x \sqrt{g} \,\nabla_\mu ( \mu \, \phi \, \frac{\epsilon^{\mu\nu\lambda\sigma}}{\sqrt{g}} A_{\nu\lambda\sigma})/6$. The membrane is charged under the $4$-form, that can jump between interior and exterior of the membrane, changing according to $\Delta F_{\mu\nu\lambda\sigma} = e \sqrt{g} \epsilon_{\mu\nu\lambda \sigma}$. In addition to the global dynamics controlled by membrane emission, in the presence of nonzero mixing $\mu \ne 0$ the $4$-form is {\it not} locally constant \cite{giavilen,ksQfin}. It depends on the scalar field $\phi$, which mixes with it and becomes {\it massive}: the $4$-form background gives inertia to the scalar's propagation, which by local Lorentz invariance translates into the scalar mass term. After the background is selected the $4$-form locks to $\phi$, breaking the shift symmetry {\it spontaneously} \cite{giavilen}. 

A representation which manifestly displays the above features follows if we integrate out the $4$-form, bearing in mind that the membrane emission can change its background value \cite{ksQfin}.  So, using the first order formalism by extending the action with the Lagrange multiplier term 
${\cal S}_q = \int d^4 x \,   \left(q/{24}\right) \, \epsilon^{\mu\nu\lambda\sigma} \, \left(
F_{\mu\nu\lambda\sigma} -  4\, \partial_{\mu} A_{\nu\lambda\sigma} \right)$ \cite{nktop,ksQfin}, completing the squares in $F_{\mu\nu\lambda\sigma}$, properly accounting for the boundary terms and integrating 
$F$-dependent terms out, we get
\begin{eqnarray}
{\cal S}_{eff} &=&\int d^4 x \sqrt{g} \left( \frac{M_{Pl}^2}{2} R -\frac12({\nabla}\phi)^2 
- \frac12 (q + \mu \phi)^2 + \right. \nonumber\\
&+&\left. \frac16 \frac{\epsilon^{\mu\nu\lambda\sigma}}{\sqrt{g}} \, A_{\nu\lambda\sigma} \, \partial_{\mu} q \right) \, .
\label{effact}
\end{eqnarray}
The boundary term (\ref{actionbrane}) depending on the membrane charge also remains, now giving the global dynamics of the Lagrange multiplier field $q$. Locally, it is an auxiliary field, since (\ref{effact}) yields 
$\partial q = 0$. The membrane term (\ref{actionbrane}) changes this, yielding a source for $\partial q$,
which jumps across the membrane by $\Delta q|_{\vec n} = e$~.

Eq.~(\ref{effact}) shows that the mixing has induced an effective potential
$V = \frac12 (q + \mu \phi)^2$, instead of the pure cosmological constant contribution $\frac12 q^2$, where the scalar field has mass $\mu$ and, for a fixed $q$, the minimum at $\phi_{min} = -q/\mu$. 
The shift symmetry $\phi \rightarrow \phi + \phi_0$ survives in the action because the variation of $\phi$ is compensated by the shift $q \rightarrow q - \mu \phi_0$. Once $q$ is fixed as a solution of the field equations, the shift symmetry is broken spontaneously.

When considering the mass of $\phi$, one has to worry about possible competing contributions from other corners of the theory. 
The presence of the shift symmetry in the action (\ref{effact}) protects the {\it massive} field  $\phi$ from radiative corrections to its mass. It implies that $\phi$ couples to other matter only derivatively, and so radiative corrections induced by such couplings won't change the mass term. On the other hand, if the axion couples to some gauge theory with the standard Chern-Simons term $\sim \frac{\phi}{f_\phi} \, Tr(G \wedge G)$, the instanton effects will break the shift symmetry down to its discrete subgoup $\phi \rightarrow \phi + n \pi f_\phi$. The resulting effective potential will contribute to the axion mass, and in fact in the standard axionic inflation models, it is this potential that one uses for driving inflation \cite{natural}. But as we noted above, this requires $f_\phi > M_{Pl}$. If on the other hand the converse holds, as is argued to be more natural in {\it UV} complete theories \cite{tometc}, this contribution to the potential may become an obstruction if it is too large. However when $f_\phi < M_{Pl}$ as long as the scale of the potential $\lambda$ obeys $\lambda^4 < \mu^2 f_\phi^2$ the instanton corrections will remain by and large negligible, merely yielding small bumps on top of the potential 
$\frac12 (q + \mu \phi)^2$ \cite{giavilen}. 

Another concern regarding the flatness of the $4$-form induced potential comes from considering corrections from higher dimension operators, omitted in (\ref{actionbulk}). By gauge symmetry of $F$ and shift symmetry of $\phi$ they can be organized as an expansion in $F^{n+2}/M^{2n}$ where $M$ is the $UV$ cutoff, e.g. the string scale. This means that the action (\ref{actionbulk}) is a good description of the system
as long as $|F| \la M^2$. Using the on-shell form for $F$, $F_{\mu\nu\lambda\sigma} = \sqrt{g} \epsilon_{\mu\nu\lambda\sigma} (q + \mu \phi)$ \cite{ksQfin}, then 
yields the constraint $\phi \la M^2/\mu$, which still allows a wide range of variation of $\phi$~\footnote{Quantitatively the same bound comes from requiring that the axion energy density does not destabilize volume moduli in the compactifications of higher-dimensional SUGRAs that yield (\ref{actionbulk}).}.
Hence if $\mu \ll M$, the description based on (\ref{actionbulk}) remains under control, keeping the potential $V = \frac12 (q + \mu \phi)^2$ flat even when $M_{Pl} \ll \phi \la M^2/\mu$. Similar issues come up from considering gravitational effects. Perturbative corrections remain small if one starts with a flat potential, since they only give terms proportional to
$\sim m^2_{min} R$ and $\sim {V_{eff}^2}/{M_{Pl}^4}$, that are tiny as long as $V < M_{Pl}^4$~\cite{andreirad}. The gravitational instanton corrections are controlled by coefficients proportional to the exponential of the instanton action~\cite{andrlenny}. When the axion decay constant is small,
$f_\phi \la 0.1\,M_{Pl}$, which as discussed above we can choose, since we do not need it for slow roll, the instanton action will be large enough to suppress nonperturbative gravitational corrections as well. 

Let us now turn to discussing the dynamics arising from this potential. As is obvious, the $4$-form charge $q$, which determines the location of the minimum, can change by the membrane emission, and so the space of axionic vacua is really a mini-landscape, much like in \cite{andreipr}-\cite{feng}. However, as we noted in \cite{ksQfin}, the mass $\mu$ may also be a landscape variable, as models given by (\ref{actionbulk}) with $\mu \ne 0$ are naturally realized by dimensional reduction of various supergravities which arise as low energy limits of string theory. In this case, the parameter $\mu$ is in fact an internal flux of a magnetic form field, and so it is quantized just like any other generic $4$-form flux, like $q$. 
If we start from $11D$ {\it SUGRA} compactified on a $7$-torus as \cite{BP}, the expressions for the fluxes are $q_i = n_i \, e_{11}/\sqrt{Z_i}$,
where $Z_i$ are the internal volumes controlled by the (stabilized) volume moduli, and  $e_{11} = 2\pi\, M_{11}^3$ is the fundamental membrane charge, normalized to the $11D$ Planck mass $M_{11}$. The volume factors for electric (i.e. $4D$ spacetime) $4$-forms are $Z_e = M_{Pl}^2/2$, while for magnetic (i.e. internal space) $4$-forms they are $Z_{m,i}= M_{Pl}^2/\left(2 M_{11}^6 V_{3,i}^2\right)$ \cite{BP}. Since $\mu$ is the charge of a magnetic $4$-form, it is quantized according to $\mu = 2\pi \, n \, V_3 M_{11}^3  \, \left(M_{11}/M_{Pl}\right)^2 \, M_{11}$~. Thus $\mu$ can change by emission of membranes in steps of $\Delta \mu \sim  V_3 M_{11}^3  \, \left(M_{11}/M_{Pl} \right)^2 \, M_{11}$, which can be quite small. If we take a simple setup where the size of compact dimensions is not much different from the string length, which may still be sufficient to suppress the nonperturbative gravitational contributions to the axion potential, $V_3 M_{11}^3 \sim {\cal O}(10)$, the quantum of mass is 
$\Delta \mu \sim  {\cal O}(10) \times \left(M_{11}/M_{Pl}\right)^2 \, M_{11}$.

This leads to a very interesting global picture of an inflating universe. Inflation will be driven by the effective cosmological term comprised of the `bare' negative cosmological constant \cite{BP} and the sum of $4$-form fluxes which do not involve axion mixings $\Lambda\left(n_i\right)$, and the axionic inflaton term 
$\frac12 (q + \mu\phi)^2$.
The `cosmological constant' term will be eventually diminished by membrane emission, yielding somewhere in the Metaverse a net tiny cosmological constant \cite{BP}, or, if there are more axions, possibly a quintessence field in slow roll \cite{ksQfin}, either one needed to dominate the universe at the present time. The effective potential driving inflation, $V_{eff} = \Lambda(n_i) + \left(q + \mu \phi\right)^2/2$~, would support scalar field fluctuations. If the scalar fluctuations are small, they would feed into the density perturbations given by \cite{chaotic}
\be
\frac{\delta \rho}{\rho} \simeq \frac{H^2}{2\pi \dot \phi} \simeq 
\frac{[\Lambda(n_i) + \frac12 (q + \mu \phi)^2]^{3/2}}{2\pi\sqrt{3} M_{Pl}^3 \mu (q + \mu \phi)} \, ,
\label{perts}
\ee
a formula valid as long as its numerical value remains below unity. On the other hand, from the inspection of this equation, at early times when the potential is dominated by the net cosmological term $\Lambda(n_i)$, the density perturbations $\delta \rho/\rho$ can be very large. Where $\delta \rho/\rho$ exceeds unity, the quantum fluctuations of the inflaton dominate over the classical ones, and the dynamics of the field $\phi$ is going to be determined by random quantum fluctuations, under whose influence the field hops around 
preparing the regions of the Metauniverse in states where $\phi$ is suspended away from its minimum. This epoch will terminate in some regions after membrane emission reduces $\Lambda(n_i)$ to below
$\frac12 (q+\mu\phi)^2$. In those regions, $q$ and $\mu$ themselves will be random variables. Once this happens, the formula (\ref{perts}) degenerates to 
$\delta \rho/\rho \simeq \left(q + \mu \phi\right)^2/\left(4\pi \sqrt{6} M_{Pl}^3 \mu\right)\,$.

Clearly, given our bounds on the maximal value of $\phi$ for which we can still use the low energy action (\ref{actionbulk}), and the estimates above, this region of the universe may still be trapped in the self-reproduction regime after the membranes have carried away $\Lambda(n_i)$. Or not -- in any case, eventually in some regions quantum effects will take the inflaton away from the self-reproduction regime.
At that point, the standard slow roll inflation will begin, creating a large inflated domain. As the inflaton background value $q/\mu + \phi$ falls below $M_{Pl}$, 
inflation will terminate, and the inflaton will begin to oscillate about the local minimum at $\phi = - q/\mu$, 
reheating this region of the universe in the process. Reheating may occur by the production of for example the gauge sector to which the inflaton may couple by $\sim \frac{\phi}{f_\phi} \, Tr(G \wedge G)$,
and subsequent thermalization of this gauge theory with the Standard Model particles. The reheating temperature would be $T_R \sim \sqrt{\Gamma_\phi M_{Pl}}$, where 
$\Gamma_\phi \sim \mu^3/f_\phi^2$ is the decay rate of $\phi$ into the gauge fields $G$. Thus, $T_R \sim \sqrt{\mu^3 M_{Pl}}/f_\phi$,
well above the temperature needed for nucleosynthesis. 

An important point which needs to be stressed here is that when $\mu$ is a random variable, so are the number of efolds which unravel during slow roll phase and the value of the amplitude of the nearly scale invariant spectrum of density perturbations, changing from one slow-roll region to another. Indeed, these quantities depend on $q$ and $\mu$, as is straightforward to calculate. Assuming that the slow roll started with the value of $\phi$ at the threshold of self-reproduction, as suggested by the global picture outlined above, they are \cite{chaotic} $N_* \simeq \sqrt{6\,\pi}\,M_{Pl}/\mu\,$, $\delta \rho/\rho\simeq 10 \,\mu/M_{Pl}\,$, where we have normalized the perturbations to their value 60 efolds before the end of inflation. Clearly, these change from one low energy universe to another (as does the reheating temperature $T_R$) but the {\it a posteriori} requirement of producing a universe which has inflated at least 60 efolds makes the dependence on $q$ very weak. Nevertheless, this may still provide one with an arena to explore anthropic reasoning further, by allowing for jumps in $\mu$ during the last stages of inflation, that could yield to inflating domains whose boundaries might still be visible. In such cases one could search for the variation of both residual curvature of cosmological spatial slices and the amplitude of density perturbations, as probed in \cite{lennymatt}. We will not delve into this interesting and important arena here. We will merely note
that the requirement that the density perturbations are of the right scale, $\delta \rho/\rho \simeq 10^{-5}$, which implies $\mu \simeq 10^{13} \, {\rm GeV}$, can be directly related to GUT scale physics if we take the inflaton to have no more than few units of the quantum of mass, $\sim {\cal O}(10) \times (\frac{M_{11}}{M_{Pl}})^2 M_{11}$, during the final stage of inflation in our region of the Metaverse. Indeed, it is easy to check that we need $M_{11} \sim 10^{15} \, {\rm GeV}$. 

To conclude, we have shown that the simplest scenario of chaotic inflation can be naturally realized in theories where axionic fields mix with $4$-forms. The resulting low energy theory yields a model with a quadratic potential generated by the mixing, and protected from higher order corrections in perturbation theory by a shift symmetry, that remains unbroken at the level of the action. The nonperturbative contributions to the potential both from field theory and from gravity may be suppressed when 
$f_\phi < M_{Pl}$, if the gauge theory to which the inflaton couples is not strong at too high a scale. The structure of the vacuum configurations is a mini-landscape, and in some regions the conditions for successful chaotic inflation will occur automatically. In them, the value of density perturbations may be a random variable, as it depends on the inflaton mass. This will occur in the theories where the effective $4D$ picture which we adopt arises after dimensional reduction, where the inflaton mass is also one of the form fluxes. In that case it can change from place to place, being decreased by membrane emission. This can be an interesting scene for testing anthropic ideas and general features of the landscape approach to cosmology. \\

{\bf \noindent Acknowledgements} We thank Shamit Kachru, Renata Kallosh, Andrei Linde, Markus Luty, Alessandro Tomasiello, Arkady Vainshtein and Jun'ichi Yokoyama for valuable discussions.
LS thanks the UC Davis HEFTI program
for hospitality during the inception of this work. The work of NK is
supported in part by the DOE Grant DE-FG03-91ER40674. 
The work of LS is partially
supported by the U.S. NSF Grant
PHY-0555304.


\end{document}